\documentclass[aps,prl,reprint,superscriptaddress]{revtex4-2}

\usepackage{amssymb}
\usepackage{amsmath}
\usepackage{epsfig}
\usepackage{epstopdf}
\usepackage{bm}
\usepackage{graphicx,epsfig}
\usepackage{mathrsfs}
\usepackage{dcolumn}
\usepackage{color}
\usepackage{natbib}
\usepackage{CJK} 
\usepackage{url}
\usepackage{upgreek}

\begin{document}

\title{Extreme subradiance from two-band Bloch oscillations in atomic arrays}

\author{Luojia Wang}
\affiliation{State Key Laboratory of Advanced Optical Communication Systems and Networks, School of Physics and	Astronomy, Shanghai Jiao Tong University, Shanghai 200240, China}

\author{Da-Wei Wang}
\affiliation{Interdisciplinary Center for Quantum Information, State Key Laboratory of Modern Optical Instrumentation, and Zhejiang Province Key Laboratory of Quantum Technology and Device, Department of Physics, Zhejiang University, Hangzhou 310027, Zhejiang Province, China}
\affiliation{CAS Center for Excellence in Topological Quantum Computation, University of Chinese Academy of Sciences, Beijing 100190, China}

\author{Luqi Yuan}
\email{yuanluqi@sjtu.edu.cn}
\affiliation{State Key Laboratory of Advanced Optical Communication Systems and Networks, School of Physics and	Astronomy, Shanghai Jiao Tong University, Shanghai 200240, China}

\author{Yaping Yang}
\email{yang\underline{ }yaping@tongji.edu.cn}
\affiliation{MOE Key Laboratory of Advanced Micro-Structured Materials, School of Physics Science and Engineering, Tongji University, Shanghai, 200092, China}

\author{Xianfeng Chen}
\email{xfchen@sjtu.edu.cn}
\affiliation{State Key Laboratory of Advanced Optical Communication Systems and Networks, School of Physics and	Astronomy, Shanghai Jiao Tong University, Shanghai 200240, China}
\affiliation{Collaborative Innovation Center of Light Manipulations and Applications, Shandong Normal University, Jinan, 250358, China}



\begin{abstract}
Atomic arrays provide an important quantum optical platform with photon-mediated dipole-dipole interactions, which can be engineered to realize key applications in quantum information processing. A major obstacle for such application is the fast decay of the excited states. By controlling two-band Bloch oscillations in an atomic array under external magnetic field, here we show that exotic subradiance can be realized and maintained at a time scale upto 12 orders of magnitude larger than the spontaneous decay time in atomic arrays with the finite size. The key finding is to show a way for preventing the wavepacket of excited states scattering into the dissipative zone inside the free space light cone, which therefore leads to the excitation staying at a subradiant state for extremely long decay time. We show that such operation can be achieved by introducing a spatially linear potential from external magnetic field in atomic arrays and then manipulating interconnected two-band Bloch oscillations along opposite directions. Our results also point out the possibility of controllable switching between superradiant and subradiant states, which leads to potential applications in quantum storage.
\end{abstract}


\maketitle

Light-matter interaction in subwavelength scale is of broad interest in quantum information processing and quantum metrology \cite{QIRMP10,QMNP11}. Such interactions in subwavelength atomic arrays \cite{Subrad1DPRL13,DDGM2DNP15,Subrad1DPRA16,StoragePRX17,Scat2DPRL17,Topo1DPRA18,StorageNJP18,NanoRMP18,Subrad1DPRL19,Bell1DPRL19,ScatNP20,Subrad1DPRL20,Topo1DPRA21,StorageSPRXQ21,Storage2DPRR22,Exp1DPRX17,Exp1DS19,Exp2DRydPRL19,Exp1DPRL20,Exp2DN20,Exp2DS21,Exp2DN21,Exp2DQSLS21,Exp2DN22,Exp2DS22} can induce long-range nonlinear dipole-dipole interactions by mediating photons in radiation modes \cite{SubradPRA10,Subrad1DPRL13,Subrad1DPRA16,SubradS23} or via long-range van der Waals interactions between Rydberg states \cite{QSRydNP20}. Further controls of atomic arrays by magnetic fields bring remarkable physical phenomena including super- and subradiant states \cite{SubradStorage1DPRA16}, photon storage and retrieval \cite{Storage1DSR15,Storage2DPRL16,Storage2DPRR20}, subradiance-protected quantum state transport \cite{Trans1DNJP19}, and many others  \cite{Topo2DPRL17,Topo2DPRA17,Topo2DBEPRA17,Topo2DQST18,Topo1DCP19,Topo2DPRL20,Scat2DPRA18,Scat2DPRA19}. These quantum information processing methods utilize subradiant quantum states with inhibited spontaneous decay and exhibit useful applications in quantum storage \cite{QMPT15}. Previous researches show that subradiant states in one-dimensional (1D) arrays of $N$ atoms typically have decay rates scaling as $N^{-3}$ with a lifetime 7 orders of magnitude greater than that of a single atom for $N\sim 200$ \cite{StoragePRX17,Subrad1DPRL19,Subrad1DPRL20}. Nevertheless, it generally desires a quantum state lasting long enough for the purpose of quantum storage \cite{QMPT15}. A challenging problem is that subradiant states can diffuse into the superradiant subspace through interaction or boundary effect, which limits their applications in quantum information processing. 

\begin{figure}[!htb]
\centering
\includegraphics[width=1\columnwidth]{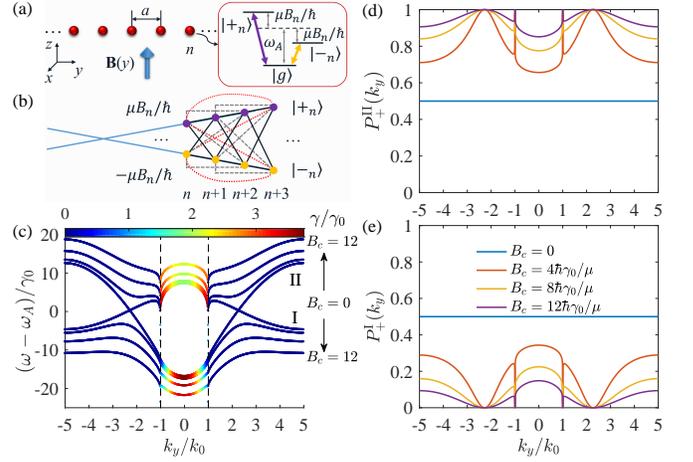}
\caption{(a) Schematic of a 1D V-type atomic array under a magnetic field $B(y)$. (b) Schematic of the corresponding lattice where the $n$th atom has non-degenerate excited states $\left|\pm_n \right\rangle$ with frequency shifted by $\pm \mu B_n/\hbar$. Blue lines indicate a linear trend of $B_n$ as in Eq. \eqref{EqB}. (c) Band structures with different constant magnetic field $B_c$ with values $B_c=0,4,8,12\hbar\gamma_0/\mu$, respectively. Black arrows indicate values of $B_c$ for corresponding bands. Decay rates of modes are color coded. Probabilities of eigenstates on (d) band II and (e) band I projected on $\left|+\right\rangle$ states for different $B_c$. Here $a=0.1\lambda$.
\label{Figarray}}
\end{figure}

Here we show that a judicious coherent control of finite-size atomic arrays can prevent subradiant states from entering the superradiant subspace and thus realize subradiance with a lifetime $10^{12}$ times longer than that of a single atom. Specifically, we consider a well-established model with atomic arrays under the magnetic field which lifts the degeneracy of excited states in each atom through the Zeeman shift (labelled as $\left|+\right\rangle$ and $\left|-\right\rangle$) \cite{Topo2DPRL17,Topo2DPRA17,Topo1DCP19}. Such systems with the infinite size can support subradiant collective excited states outside of the dissipative zone in the momentum space \cite{StoragePRX17,Subrad1DPRL19,Subrad1DPRL20}. However, for the same system with a finite size, such excitation  inevitably spreads to boundaries and the wavepacket gets scattered into the dissipative zone. To overcome such obstacle, we instead apply a spatially linear magnetic field which brings linear potentials with opposite slopes on $\left|+\right\rangle$ and $\left|-\right\rangle$ states [see Fig. \ref{Figarray}(b)]. Similar 1D lattices under linear potentials support Bloch oscillations \cite{PTBOPRL09,RMZakNP13,ExpHaldaneBON14,ZeemanBOPRA15,FlatBOPRL16,ZeemanBOPRL16,LadderBOPRA17,LadderBOPLA19,SOCBOPRA19,PTN12,RingBOOpt16,EITBOO17}. Fundamentally different from these previous works, Bloch oscillations in atomic arrays under linear magnetic field support two-band Bloch oscillations with inter-band interactions such that the wavepacket of excited states oscillates on two bands along opposite directions alternatively. This unique picture can force the wavepacket spatially localized within a few atoms while preventing the excitation entering into the dissipative zone.  Hence it supports subradient oscillations surviving extremely long time with finite atoms. We also show the way of switching between superradiant and subradiant oscillations by tuning the external magnetic field, which provides the opportunity for reading out information from subradiant quantum states \cite{ControlPRL15,ControlExpPRL20,BatteryPRL18,BatteryPRL19}.

\textbf{Model}. The developments of state-of-art technologies in various experimental platforms such as trapped neutral atoms at subwavelength scale \cite{Exp1DPRX17,Exp3DOLN18,Exp2DPRL19,Exp2DRydPRL19,Exp1DPRL20,Exp2DN20,Exp2DS16,Exp1DS16,Exp2DPRX18,Exp1DS19,Exp2DS21,Exp2DN21,Exp2DQSLS21,Exp2DN22,Exp2DS22} pave the way for exploring exotic phenomena in atomic arrays. We study such arrays of atoms that are equally spaced at a distance $a$ along the $y$ axis [$y_n=na$ for the $n$-th atom in Fig. \ref{Figarray}(a)]. Each atom has a ground state $\left|g \right\rangle$ and degenerate excites states. We apply an off-plane magnetic field $B(y)$ in $z$ direction to lift the degeneracy of two excited states $\left|\pm \right\rangle$. Under the single-excitation limit, the dynamics of the wave packet of excited states is described by the non-Hermitian Hamiltonian after integrating out photonic modes \cite{Topo2DPRL17,Topo2DPRA17,Topo1DCP19}:
\begin{equation}
\begin{aligned}
H_{\mathrm{eff}}=&\hbar\sum_{n=1}^{N}\sum_{\alpha=\pm}\left(\omega_{A}-i\frac{\gamma_{0}}{2}\right)\left|\alpha_{n}\right\rangle \left\langle \alpha_{n}\right|\\
&+\sum_{n=1}^{N}\mu B_n\left(\left|+_n\right\rangle \left\langle +_n\right|-\left|-_n\right\rangle \left\langle -_n\right|\right)\\
&+\frac{3\pi\hbar\gamma_{0}c}{\omega_{A}}\sum_{n\neq m}\sum_{\alpha,\beta=\pm}G_{\alpha\beta}\left(y_n-y_m\right)\left|\alpha_n\right\rangle \left\langle \beta_m\right|,
\end{aligned}\label{EqHeff}
\end{equation}
where $\omega _A$ is the atomic transition frequency, $\gamma_0$ is the atomic decay rate in the free space, $\mu B_n/\hbar$ is the Zeeman shift for the $n$th atom with the magnetic moment $\mu$, and $G_{\alpha\beta}\left(y_n-y_m\right)$ is the free-space dyadic Green's function describing the electric
field at $y_n$ emitted by the atom located at $y_m$. With the relation $\left|\pm \right\rangle=\mp\left(\left|x \right\rangle \pm i\left|y \right \rangle \right) / \sqrt{2}$, one obtains $G_{++}=G_{--}=-e^{ik_0r} \left(k_0^2r^2-ik_0r+1\right)/\left( 8\pi k_0^2r^3\right) $ and $G_{+-}=G_{-+}=e^{ik_0r} \left(k_0^2r^2+3ik_0r-3\right)/\left( 8\pi k_0^2r^3\right) $, where $r=\left| y_n-y_m\right| $ and $k_0=\omega_A/c=2\pi/\lambda$ with $\lambda$ being the wavelength and $c$ being the vacuum speed of light \cite{Topo2DPRA17,Topo1DCP19}. Eq. \eqref{EqHeff} shows a tight-binding lattice model where excitations on sites at two arms ($\left|\pm \right\rangle$) have on-site potentials $V_{\pm,n} \equiv \pm \mu B_n$ and are connected by complex and long-range photon-mediated hoppings.

We first study the band structure of infinite atomic arrays with constant magnetic field $B_c$ \cite{Topo2DPRL17,Topo2DPRA17}. We choose $a=0.1\lambda$, which brings the first Brillouin zone $k_y\in \left[-5k_0,5k_0\right]$. We plot band structures in $k_y$-space for different $B_c$ in Fig. \ref{Figarray}(c). For each $B_c$, there are two bands, and it exhibits large collective decay, i.e., the dissipative zone in the free-space light cone ($\left|k_y\right|<k_0$). Outside of the dissipative zone, subradiant states are supported. For $B_c=0$, there are two degenerate points near $k_y=\pm 2.3k_0$, while two bands are separated in the entire $k_y$-space for other $B_c$. We denote the lower (upper) band as band I (II) and define $P^{\mathrm{I(II)}}_\pm$ as the probability intensity of an eigenstate on band I (II) projected onto the arm of $\left|\pm\right\rangle$ states. $P_+^{\mathrm{II}}$ and $P_+^{\mathrm{I}}$ are shown in Figs. \ref{Figarray}(d) and \ref{Figarray}(e), respectively. One notes that $P^{\mathrm{I(II)}}_+ + P^{\mathrm{I(II)}}_-=1$. When there is no magnetic field, $P^{\mathrm{I}}_\pm = P^{\mathrm{II}}_\pm=0.5$ in the entire $k_y$-space, indicating that probabilities for exciting $\left|\pm\right\rangle$ states are equivalent due to the degeneracy of $\left|\pm\right\rangle$. However, for $m \neq 0$, it shows asymmetric distributions of $P^{\mathrm{I(II)}}_\pm$. In particular, for $B_c>0$, on band I (II), we can see that $P^{\mathrm{I}}_+ < 0.5$ ($P^{\mathrm{II}}_+ > 0.5$), meaning that less (more) density of state is located on the arm of $\left|+\right\rangle$ states. Distributions of $P^{\mathrm{I(II)}}_\pm$ are highly symmetric because the influence of $\left|\pm\right\rangle$ is symmetric under the positive/negative magnetic field. Moreover, we find that, for $B_c>0$, at $k_y=\pm 2.3 k_0$, it shows $P^{\mathrm{I}}_+ =0$ ($P^{\mathrm{II}}_+ =1$) on band I (II). We note that band structures in Fig. \ref{Figarray} are calculated by simply summing the Green's function in real space over all lattice sites, which artificially induces the inaccurate eigenvalues at $k_y$ near $\pm k_0$ in the free-space light cone due to slow convergence. However, this inaccuracy in band structures does not affect the later analysis throughout this paper. Also, if desired, it can be overcome by summations of Green's function in momentum space with an appropriate regularizing method \cite{Topo2DPRL17,Topo2DPRA17}.

We next introduce the simulation method for studying the dynamics of the excitation wavepacket in atomic arrays. We use the Schr\"{o}dinger equation $\textbf{d}\left|\Psi\left( t\right) \right\rangle/\textbf{d}t = - i H_{\textrm{eff}} \left|\Psi\left(t\right)\right\rangle/\hbar$ and the excitation wavepacket of the atomic array
$\left|\Psi\left( t\right) \right\rangle =\sum_n [ C_{+,n} \left(t\right) \left|+_n\right\rangle + C_{-,n} \left(t\right) \left|-_n\right\rangle ] e^{-i\omega_A t}$,
where $\left| C_{\pm,n}\right|^2$ gives the excitation probability of the $\left|\pm\right\rangle$ state in the $n$-th atom. 201 atoms ($n=-100,\ldots,0,\ldots,100$) are considered in arrays. We assume that the wavepacket of the system is initially prepared at a superposition state on two arms following
$C_{\pm,n} \left(0\right) = \psi_{\pm} \exp\left[  ik_can-\left(n-n_c \right)^2/200\right]$, where $n_c$ is the spatial center of the initial excitation, $k_c$ is the initial momentum, and $\psi_{\pm}$ gives the initial ratio of excitation amplitudes on two arms of $\left|\pm\right\rangle$ states. Such an initial state in the weak excitation limit can be pumped with phase-controlled schemes \cite{SubradStorage1DPRA16,Storage1DSR15,ControlPRL15,ControlPRL20} or by applying a spatial modulation of the atomic detuning \cite{Storage2DPRR22}.

\begin{figure}[!htbp]
\centering
\includegraphics[width=1\columnwidth]{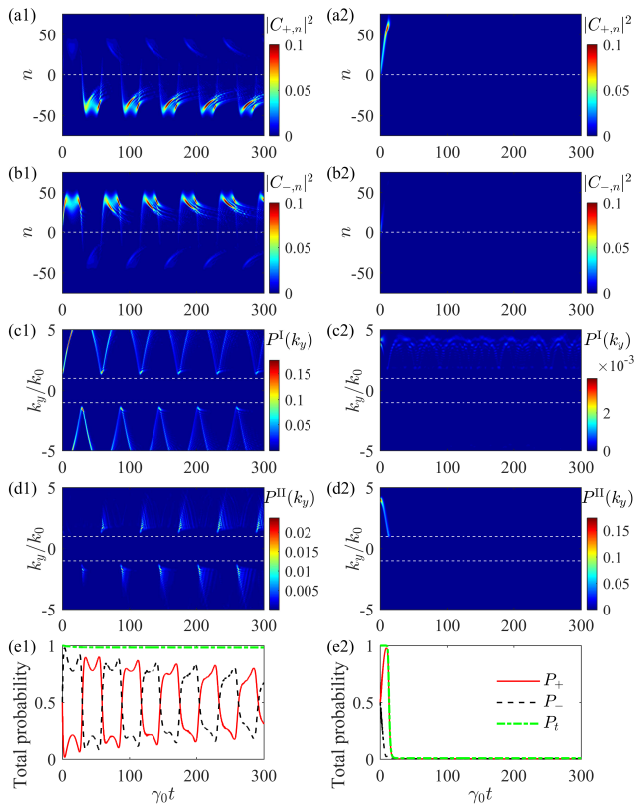}
\caption{Bloch oscillations for Gaussian excitations initially centered at $n_c=0$ with  $k_c=1.5k_0$ on band I (left) and $k_c=4k_0$ on band II (right), shown by temporal evolution for excitation probabilities of (a1),(a2) $\left| C_{+,n}\right|^2 $, (b1),(b2) $\left| C_{-,n}\right|^2 $, (c1),(c2) $P^{\mathrm{I}} \left(k_y\right)$, (d1),(d2) $P^{\mathrm{II}} \left(k_y\right)$, (e1),(e2) $P_+$, $P_-$, and $P_t$. Here $a=0.1\lambda$ and $\mu B_0/\hbar=0.2\gamma_0$.}
\label{FigkI1.5}
\end{figure}

\textbf{Results}. For finite atomic arrays, the key ingredient for maintaining long-standing subradiate state is to prevent the wavepacket entering into the dissipative zone either by oscillations or scattering due to the boundary effects. One might notice that the well-established Bloch oscillation may lead the wavepacket oscillating within a finite spatial region and propagating unidirectionaly in the $k_y$-space \cite{BONJP04,ZeemanBOPRA15,LadderBOPRA17}, so by frequently altering the direction of the constant force, it is possible to accomplish the task for avoiding entering the dissipative zone. Nevertheless, here we show that one can realize such task in atomic arrays under a spatially linear but temporally constant magnetic field. In particular, we consider the magnetic field
\begin{equation}\label{EqB}
B_n = nB_0,
\end{equation}
where $B_0$ is a constant. Eq. \eqref{EqB} gives $V_{\pm,n} =\pm n\mu B_0 $, which leads to the effective constant force $F_\pm =\mp \mu B_0/a$ in opposite directions on the two excited states. Two effective electric fields at opposite directions on two separated arms in the lattice bring Bloch oscillations on two arms exhibiting symmetric patterns. However, once the two arms are connected [by terms including $G_{\pm\mp}$ in Eq. \eqref{EqHeff}], the dynamics of the Bloch-oscillation wavepackets on two arms are influenced by each other. We take $B_0=0.2\hbar\gamma_0/\mu$ in simulations.

To see the long-standing subradiant state, we excite the wavepacket centered at $n=0$, i.e., $n_c=0$, with $\psi_+=\psi_-=0.168$ and $k_c=1.5k_0$, which satisfies $\sum_n\left[ \left|C_{+,n}(0)\right|^2 + \left|C_{-,n}(0)\right|^2 \right]=1$. The evolutions in the simulation are plotted for the time upto $10 T_B$, where $T_B=2\pi\hbar/\mu B_0=10\pi/\gamma_0$. Figs. \ref{FigkI1.5}(a1) and \ref{FigkI1.5}(b1) depict the dynamics of $\left| C_{+,n} \left(t\right) \right|^2$ and $\left| C_{-,n} \left(t\right) \right|^2$ respectively and one sees a sustained Bloch oscillation pattern with the shape of the wavepacket deforming gradually. The excitation of $\left| +\right\rangle $ ($\left| -\right\rangle$) states in atomic arrays is mainly located in the finite region with $y<0$ ($y>0$) so there is no boundary effects due to finite atoms. 

We then perform fast Fourier transform (FFT) on simulation results $C_{\pm,n}\left(t\right)$, and calculate the projection of the excited wavepacket in $k_y$-space onto the two bands (excitation probabilities $P^{\mathrm{I}} \left(k_y, t\right)$ and $P^{\mathrm{II}} \left(k_y, t\right)$, respectively). Note that the Bloch band picture is valid only with constant magnetic field, the linear magnetic field and resulting constant force here brings perturbation so we can calculate the projection by taking into account of uniform potentials at the value of $V_{\pm,n}=\pm \bar{n} \mu B_0$, where $\bar{n}\equiv \left\langle \Psi \right| \sum_n \left(\left|+_n\right\rangle n \left\langle +_n\right|+\left|-_n\right\rangle n \left\langle -_n\right|\right) \left| \Psi \right\rangle $ is the mean position of the wavepacket. This analysis in the Bloch band picture with uniform on-site potentials reveals dynamical features different from a conventional Bloch oscillation. Figs. \ref{FigkI1.5}(c1) and \ref{FigkI1.5}(d1) show the evolution of excitation probabilities of band I and band II, $P^{\mathrm{I}}$ and $P^{\mathrm{II}}$, at each $k_y$. The choice of $\psi_\pm$ makes the initial excitation on band I, and throughout the evolution, $P^{\mathrm{II}}$ is about one order of magnitude smaller than $P^{\mathrm{I}}$. Remarkably, one sees that $P^{\mathrm{I}}$ doesn't evolve uni-directionally on $k_y$ with time, which gives the fundamental difference from conventional Bloch oscillations. The evolution direction of $P^{\mathrm{I}}$ reverses every time (referred as the \textit{reverse time}) before it enters the free-space light cone in the $k_y$ axis [dash lines in Figs. \ref{FigkI1.5}(c1) and \ref{FigkI1.5}(d1)], which leads to the important consequence of avoiding large collective decay of the excited wavepacket. We further show the total excitation probabilities of the $\left| +\right\rangle $ state and the $\left| -\right\rangle $ state ($P_+$, $P_-$) in the real space and the total excitation probability $P_{t} = P_+ + P_-$ in Fig. \ref{FigkI1.5}(e1). We find that times that $P_+=P_-$ are exactly the same as reverse times because the expectation value of the effective electric force on the excited wavepacket on two arms, 
$\left\langle F\right\rangle  \equiv \frac{\mu B_0}{a}\sum_n \left\langle \Psi\right| \left( \left|+_n\right\rangle \left\langle +_n\right|-\left|-_n\right\rangle \left\langle -_n\right| \right)  \left| \Psi \right\rangle=\frac{\mu B_0}{a}\sum_n \left(P_+ - P_-\right)$,
becomes zero and changes its direction subsequently. In other words, when the center of the wavepacket arrives at $n=0$, $P_+ \approx P_-$ and hence $\left\langle F\right\rangle$ changes its direction, resulting in the reverse of $P^{\mathrm{I}}\left(k_y \right)$ in the momentum space. Moreover, $P_t$ exhibits no decay in this dissipative atomic system, which is actually lasting $\sim 10^8 \gamma_0^{-1}$ in a same simulation with a much longer evolution time. Therefore, as long as the excited wavepacket in atomic arrays doesn't enter the dissipative zone inside the free-space light cone, the life time of the excitation gives the subradiant feature, which is much longer than the free-space decay time. Such extreme subradiant oscillations can be optimized by choosing initial excitations. In particular, we find that $P_t$ can last more than $\sim 10^{12} \gamma_0^{-1}$ for an optimum initial state centered at $k_c=3k_0$ on band I \cite{supp}.

As one may notice that the initial choice of the excited wavepacket determines how the wavepacket evolves in $k_y$-space and hence how the wavepacket decays. For example, if we choose $\psi_+=0.168$, $\psi_-=-0.168$, and $k_c=4k_0$ to excite band I or $\psi_+=0.168$, $\psi_-=-0.168$, and $k_c=2k_0$ to excite band II, we find the total probability of wavepacket undergoes no decay even though the evolution patterns are different (see the supplementary for details \cite{supp}). However, in the case of $\psi_+=\psi_-=0.168$ and $k_c=4k_0$, with band II being excited, the dramatically different dynamics exhibits, as shown in Figs. \ref{FigkI1.5}(a2)--\ref{FigkI1.5}(e2). One notes that the excitation undergoes a rapid decay after the partial oscillation during the initial time duration $\sim11 \gamma_0^{-1}$. The fundamental difference here is that the excitation probability $P^\mathrm{II}$ moves into the free-space light cone along the $k_y$ axis before $\left\langle F\right\rangle$ changes its direction. This result leads to a very fast decay within $\sim8 \gamma_0^{-1}$ for the total probability $P_t$ dropping to $\sim0.02$. In real space, the excitation on the $\left| -\right\rangle $ arm decreases fast, while the excitation on the $\left| +\right\rangle $ arm increases initially with a oscillation towards the positive $y$-axis but decays rapidly afterwards.

\begin{figure}[!htbp]
	\centering
	\includegraphics[width=1\columnwidth]{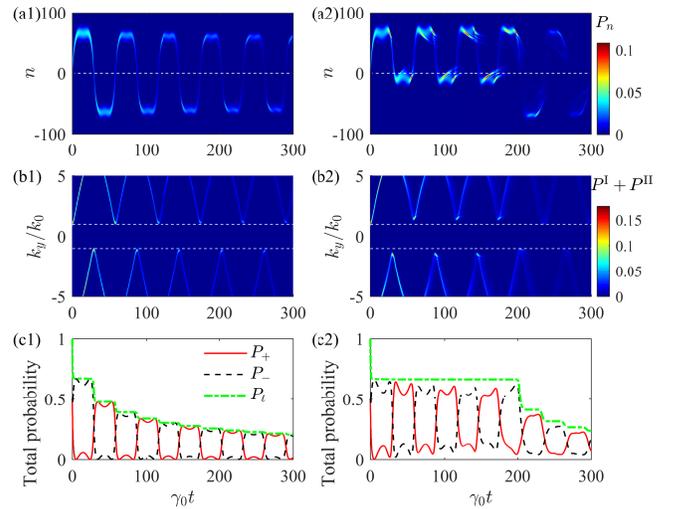}
	\caption{Bloch oscillations for Gaussian excitations initially centered at $n_c=0$ and $k_c=k_0$ on band I with a static magnetic field (left) and a controllable magnetic field (right), whose zero point is shifted to $y=30a$ over a time period from 0 to $20\gamma_0^{-1}$ and back to $y=0$ over a time period from $180 \gamma_0^{-1}$ to $200 \gamma_0^{-1}$, shown by temporal evolution of (a1),(a2) $P_{n}$, (b1),(b2) $P^{\mathrm{I}}+P^{\mathrm{II}} $, (c1),(c2) $P_+$, $P_-$, and $P_t$. Other parameters are the same as those in Fig. \ref{FigkI1.5}.} \label{FigkI1}
\end{figure}

The two-band Bloch oscillation in atomic arrays under linear magnetic field brings unique opportunity for manipulating the decay of the excited wavepacket. For instance, we can control the collective decay exhibiting quantization-like decay with stable plateaus as shown in Fig. \ref{FigkI1}(c1). To achieve it, we choose $\psi_+=\psi_-=0.168$ and $k_c=k_0$, so the wavepacket on band I centered at $k_0$ is initially excited. The simulation results are summarized in Figs. \ref{FigkI1}(a1)--\ref{FigkI1}(c1). The wavepacket experiences decay at $t\sim0$ because the initial wavepacket is partially inside the free-space light cone. However, the wavepacket initially moves towards the positive direction in the $k_y$ axis [see $P^{\mathrm{I}} \left(k_y\right) +P^{\mathrm{II}} \left(k_y\right) $ in Fig. \ref{FigkI1}(b1)], so $P_t$ ends in a stable plateau at $t\sim2\gamma_0^{-1}$ with a value $\sim0.67$. At $t\sim 27\gamma_0^{-1}$, the excited wavepacket reaches at $k_y=-k_0$ and a reverse happens due to the change of $\left\langle F\right\rangle$, which results in a rapid drop of $P_t$ to $\sim0.48$. Periodic oscillations exhibit, together with the oscillations of $P_n=\left| C_{+,n}  \right|^2 +\left| C_{-,n} \right|^2$ in the real space. The total excitation probability $P_t$ therefore shows the overall decay tendency with periodic plateaus [see Fig. \ref{FigkI1}(c1)]. Initial excitations inside the free-space light cone can exhibit similar oscillations but quicker decay \cite{supp}. Especially for an initial wavepacket at $k_c=-0.5k_0$ on band II, $P_t$ decays to $\sim 0.02$ within $t\sim 0.85\gamma_0^{-1}$, or a collective superradiant decay rate of $\sim 4.7\gamma_0$. Moreover, excitations with $n_c \neq 0$ can bring different Block oscillation phenomena because the central momenta at reverse times are changed during the evolution of the wavepacket \cite{supp}.

Taking advantage of the relation between reverse times and the atomic position in the linear magnetic field, one can control the momentum of the wavepacket at the reverse time. We give an example in Figs. \ref{FigkI1}(a2)--\ref{FigkI1}(c2), where the spatial distribution of the linear magnetic field relative to the atomic position is tuned to control the evolution of the wavepacket that is initially excited at $k_0$ on band I [the same as that in Figs. \ref{FigkI1}(a1)--\ref{FigkI1}(c1)]. In the numerical simulation, we tune the zero point of the magnetic field from $y=0$ to $y=30a$ over a time period from 0 to $20\gamma_0^{-1}$. As a result, after $P_t$ drops to the first plateau, the wavepacket oscillates around the new center $y=30a$ [Fig. \ref{FigkI1}(a2)] and reverses at $\sim \pm1.4 k_0$ in the momentum space to maintain subradiant [Fig. \ref{FigkI1}(b2)]. The excitation can also be driven to the dissipative zone by moving the zero point of the magnetic field back to $y=0$ over a time period from $180\gamma_0^{-1}$ to $200\gamma_0^{-1}$ and decay radiatively shown by descending plateaus in the total excitation probability.

\textbf{Discussions}. The proposed atomic arrays can be realized in various experimental systems such as neutral atoms trapped in optical lattices \cite{Exp1DPRX17,Exp3DOLN18,Exp2DPRL19,Exp2DRydPRL19,Exp1DPRL20,Exp2DN20} or optical tweezers \cite{Exp2DS16,Exp1DS16,Exp2DPRX18,Exp1DS19,Exp2DS21,Exp2DN21,Exp2DQSLS21,Exp2DN22,Exp2DS22}, atomlike color defects in diamond nanophotonic devices \cite{SiVS16,NVNRM18}, and excitons in atomically thin semiconductors \cite{TMDCNL20}.
On the other hand, to possibly decrease experimental challenges, one can increase the interatomic distance, which might limit the performance of Bloch oscillations from eigenstates properties and dispersion of energy bands \cite{supp}.

In summary, we have explored single-excitation dynamics in 1D atomic arrays under a linear magnetic field. Resulting two-band Bloch oscillations are studied. We find the evolution of the wavepacket in the momentum space depends on the expectation value of the effective electric force $\left\langle F\right\rangle $, which is determined by local properties of the band structure and thus influenced by the position of the wavepacket in the real space. Compared to the conventional Bloch oscillations, the evolution direction of the momentum reverses as long as the wavepacket passes the position where $\left\langle F\right\rangle $ changes its sign. Taking advantage of this unique property, we show the capability of generating exotic subradient oscillations that lasts upto 12 orders of magnitude greater than the atomic spontaneous decay time. Our study therefore points towards fundamental opportunities in switching quantum states between superradiant and subradiant states and realizing extreme subradiant quantum state that could be useful in the important application for the quantum storage.

\section*{Acknowledgments}
The research is supported by National Natural Science Foundation of China (12204304, 12122407, 11974245, and 12192252) and Shanghai Municipal Science and Technology Major Project (2019SHZDZX01-ZX06). L.Y. thanks the sponsorship from Yangyang Development Fund and the support from the Program for Professor of Special Appointment (Eastern Scholar) at Shanghai Institutions of Higher Learning.



\end{document}